 \documentclass[final,3p,times,authoryear]{elsarticle}

\usepackage{amssymb}
\usepackage{graphicx,amsmath}
\usepackage{ifthen}
\usepackage{endnotes}
\usepackage{hyperref}
\hypersetup{colorlinks=true,citecolor=blue}
\usepackage{sidecap}
\usepackage{breakurl}

\def\beq{\begin{equation}}
\def\eeq{\end{equation}}
\def\bea{\begin{eqnarray}}
\def\eea{\end{eqnarray}}
\def\nn{\nonumber}

\biboptions{round,comma,sort&compress}

\journal{the Journal of Evolutionary Economics}

\begin{document}

\begin{frontmatter}

\title{An age structured demographic theory of technological change}

\author{Jean-Fran\c{c}ois Mercure }
\ead{jm801@cam.ac.uk}
\address{Cambridge Centre for Climate Change Mitigation Research (4CMR), Department of Land Economy, University of Cambridge, 19 Silver Street, Cambridge, CB3 1EP, United Kingdom}

\begin{abstract}

At the heart of technology transitions lie complex processes of social and industrial dynamics. The quantitative study of sustainability transitions requires modelling work, which necessitates a theory of technology substitution. Many, if not most, contemporary modelling approaches for future technology pathways overlook most aspects of transitions theory, for instance dimensions of heterogenous investor choices, dynamic rates of diffusion and the profile of transitions. A significant body of literature however exists that demonstrates how transitions follow \emph{S}-shaped diffusion curves or Lotka-Volterra systems of equations. This framework is used \emph{ex-post} since timescales can only be reliably obtained in cases where the transitions have already occurred, precluding its use for studying cases of interest where nascent innovations in protective niches await favourable conditions for their diffusion. In principle, scaling parameters of transitions can, however, be derived from knowledge of industrial dynamics, technology turnover rates and technology characteristics. In this context, this paper presents a theory framework for evaluating the parameterisation of \emph{S}-shaped diffusion curves for use in simulation models of technology transitions without the involvement of historical data fitting, making use of standard demography theory applied to technology at the unit level. The classic Lotka-Volterra competition system emerges from first principles from demography theory, its timescales explained in terms of technology lifetimes and industrial dynamics. The theory is placed in the context of the multi-level perspective on technology transitions, where innovation and the diffusion of new socio-technical regimes take a prominent place, as well as discrete choice theory, the primary theoretical framework for introducing agent diversity.

\end{abstract}


\end{frontmatter}


\section{Introduction}

Socio-technical regime transitions are notoriously complex to model and understand quantitatively, but such an understanding may be crucial for anticipating and informing the planning of sustainability transitions. Socio-technical systems play important \emph{societal functions} \citep[][]{Geels2005,Geels2002}, and these services and their demand are in a continuous evolution. Meanwhile, the evolution of technology generates unforeseen opportunities to society that enable the creation of activities that did not exist previously, producing a complex interaction between technology, society and the economy, generating economic development through Schumpeter's widely discussed but not well understood process of `Creative Destruction' (\citealt{Schumpeter1934, Schumpeter1942,Schumpeter1939}; see also \citealt{Nelson1982}). Technological change occurs through a gradual process of technology substitutions which stems from a continuous stream of decision-making performed by a myriad of actors involved in the operation of technology or the consumption of the services it generates \citep{Grubler1998,Grubler1999}. This spans from the power sector, transport, communications and information technologies, to heating, cooling and lighting equipment and so on. In other words, technological change occurs in sectors performing societal functions where generation technologies or socio-technical regimes are not unique and competition occurs. Change in such sectors occurs through the choices of consumers or investors facing various alternatives and incomplete information, and these decisions are based, in a context of \emph{bounded rationality}, on diverse sets of considerations and constraints \citep{Nelson1982}.

The process of technological change is not currently well described by any generally accepted theory.\footnote{Observing for instance the stark contrast between approaches by \cite{Nordhaus2010} (exogenous technology trends), \cite{MESSAGE} (cost-optimisation), \cite{TIMER} (elasticities of substitution), \cite{Johansen1959,Boucekkine2011} (neoclassical vintage capital theory), \cite{Grubler1998} (empirical technology dynamics), \cite{Safarzynska2012} (evolutionary dynamics), \cite{Silverberg1988b} (self-organised systems), \cite{Malerba1999} (innovation dynamics within firms), \cite{Geels2002} (socio-technical regimes).} However, a significant and well known body of empirical literature exists that consistently describes the process of technology substitutions through gradual $S$-shaped curves (e.g. \citealt{Mansfield1961, Fisher1971, Wilson2009}; see reviews by \citealt{Grubler1998, Grubler1999}). As opposed to neoclassical technology vintage theory where capital vintages have optimal lifetimes and are treated with a reversible equilibrium theory (i.e. not path-dependent, \cite{Boucekkine2004,Johansen1959,Solow1966}, for a review see \cite{Boucekkine2011}), $S$-shaped curves suggest to adopt an approach where time and complex dynamics take a prominent role (e.g. the contagion model of \cite{Mansfield1961}). As we show here, attractive parallels with mathematical theories of population dynamics in biology, grounded in the understanding of the processes of birth and death of biological individuals (humans, animals, cells, etc), are more than simple analogies \citep[e.g.][]{Metcalfe2004,Silverberg1988b}. 

The best known such parallel is to use the Lotka-Volterra system of population growth equations of competing species in ecosystems for the competition of technologies in markets \citep[for an explanation and history see][]{Andersen1994}, or equivalently, the replicator dynamics equation of evolutionary game theory \citep{Hofbauer1998} applied to social sciences. While this idea has strong support in the field of evolutionary economics \citep{Saviotti1995, Safarzynska2010}, it also makes intuitive sense to perceive competing technologies (or even socio-technical systems) in the marketplace similarly to competing species in ecosystems (or even competing sub-ecosystems and food chains). The parallel has been brought further with the development of evolutionary game theory (for a review, see \citealt{Hodgson2012}), the pioneers of which were acutely aware of the strong analogy that could be drawn between the mathematics of the evolution of genotype frequencies and their selection in a population in biology, and the process of innovation and technology diffusion in economics. In addition to providing a definition to the concept of bounded rationality, this strand of literature demonstrates that the parallel, although described with yet insufficient precision, is more than just intuitive \citep{Metcalfe2004,Metcalfe2008}. As we show here, the missing link lies in the realm of technology selection and demography.

The description of technological change or technology evolution following parallels with biology currently remains in the conceptual and theoretical domain \citep[for a review, see][]{Safarzynska2010} or in stylised form (e.g. `history-friendly models' of \citealt{Malerba1999}; recombinant models of \citealt{Safarzynska2012}; or the replicator dynamics of \cite{Saviotti1995}). They are not quite adapted to actual quantitative applications such as modelling the supply of particular goods or services, technology mixes or the economic and environmental impacts that these may have.  Meanwhile, \cite{Geels2002}, using the multi-level perspective, describes the diffusion of socio-technical systems as much more complex than simple substitution events represented by a set of coupled differential equations, involving niches, early uncoordinated innovations and transformations in the social context, seemingly precluding any modelling attempts at all. Despite this, it is remarkable that diffusion processes have been observed in a myriad of contexts to follow a very simple ordering principle,\footnote{In the sense of complxity science, e.g. \cite{Anderson1972,Arrow1995}.} logistic curves or the more general Lotka-Volterra system of equations \citep[][and many more]{Marchetti1978, Fisher1971, Sharif1976, Nakicenovic1986, Wilson2009, Wilson2012b, Farrell1993, Lakka2012}, and that such simple patterns \emph{emerge} from the underlying complexity. 

In order to maintain a quantitative perspective in a computational model, the analysis can be restricted to the selection and diffusion component. As opposed to a fully evolutionary theory, this excludes the early erratic innovation process, assuming that new but established technologies permeate the landscape in dormant niches that could wake up, diffuse and potentially dominate given the right selection environment, for instance with targeted policy. From then onwards, the  diffusion process, gaining momentum, becomes firmer and simpler to project quantitatively. Although the quantitative prediction of technology diffusion is inherently highly uncertain, in parts due to the actual evolutionary nature of technological change, it is nevertheless a highly worthwhile venture to undertake, particularly for instance in the climate change mitigation context, in which the description of technological change is crucial in order to project energy consumption, greenhouse gas emissions and their related environmental impacts (e.g. in power generation, transport, industry). As we show elsewhere \citep{Mercure2014}, this approach offers a significant improvement over current optimisation approaches where technological change has no clear theoretical underpinning.

While concepts of technology diffusion provide insights on the key dynamics involved in transitions, they have not been used significantly in the modelling literature beyond the \emph{ex-post} empirical description of \emph{past} data, using the observed \emph{pattern}, the logistic curve \citep[][and many more]{Marchetti1978, Fisher1971, Sharif1976, Nakicenovic1986, Wilson2009, Wilson2012b}. Despite the fact that innovation may be the primary driver of economic growth \citep[][]{Nelson1982,Schumpeter1934,Schumpeter1939}, the process of technology diffusion has yet to be even considered in large scale mainstream models such as those for energy systems modelling and related energy policy analysis. These predominantly use representative agent cost-optimisation algorithms \citep[e.g. TIMES/MARKAL,][]{ETSAP} as a descriptive mechanism, which has no theoretical or empirical grounding.\footnote{No evidence points to cost-optimisation behaviour by agents, i.e. firms \citep{Nelson1982} or consumers \citep{Douglas1979}, including at an aggregate level \citep{Keen2011}, and no theory satisfactorily proves that an `average' representative agent can correctly reproduce the aggregate behaviour of an underlying diverse set of agents, including neoclassical theory \citep{Keen2011}.}
Indeed, this current lack of a representation of empirical dynamics is partly due to the fact that, while empirical diffusion measurements suggest a system for forecasting technology or market evolution, such projections would rely on measured time scaling parameters, which can be reliably measured only precisely in cases of older technologies where transitions have already occurred. Effectively, by the non-linear nature of the problem itself, obtaining such time scaling parameters for new technologies for which forecasting would be critically important cannot be reliably done based on the small amounts of available data.\footnote{E.g. fitting logistic curves requires data that spans at least beyond the inflexion point.} 

I thus ask the question, is it possible to use empirically known technology dynamics to forecast technology? If so, how can it be parameterised? As argued above, important scientific gains could be generated if new insight could be found on how to obtain these parameters through other means than the empirical fitting of diffusion data, requiring to establish a quantitative theory to understand their nature. These parameters are \emph{timescales}, and this suggests that their meaning is associated to the use, the building and the scrapping of technology in \emph{time} at the unit level, hinting to the use of demography theory applied to technology. Previous work has shown that the use of the Lotka-Volterra equation system can be made convenient, even mainstreamed, with the creation of the `Future Technology Transformations' family of computational models \citep{Mercure2012}, which enables to explore the impact and dynamics of policy instruments on choices of diverse agents, with real-world data. This model is based on the theory presented here. It was recently used to evaluate climate change impacts of combinations of policy instruments in 21 countries covering the World by integrating it to macroeconometric and climate modelling frameworks \citep{Mercure2014}, and currenlty runs under a resolution of 54 countries and 24 technologies \citep{E3MEManual}.

The goal of this paper is thus to derive a parameterisation method from a detailed theory. I first frame the problem by using an example of empirical data, and place it in context within its appropriate theoretical framework, transitions theory (section~\ref{sect:Framing}). Second, I derive from first principles components of a quantitative theory of technological change, based on survival (or demographic) analysis explaining the origin of the timescales of change (section~\ref{sect:model}). Third, by invoking theoretical concepts of technology choice mostly based on discrete choice theory, I combine the components to demonstrate how a Lotka-Volterra system (or replicator dynamics) can be derived from demography theory from first principles (section~\ref{sect:model2}). Finally, I interpret this theory by demonstrating the origin of the the scaling parameters of the Lotka-Volterra system, and point to how these can be used in real models of technology that could potentially replace with reasonable ease incumbent cost-optimisation models (section~\ref{sect:interp}).


\section{Framing the problem and putting it into theoretical context \label{sect:Framing}}
\subsection{The Lotka-Volterra equation for empirical technology transitions}

\begin{figure}[t]
		\begin{center}
			\includegraphics[width=1\columnwidth]{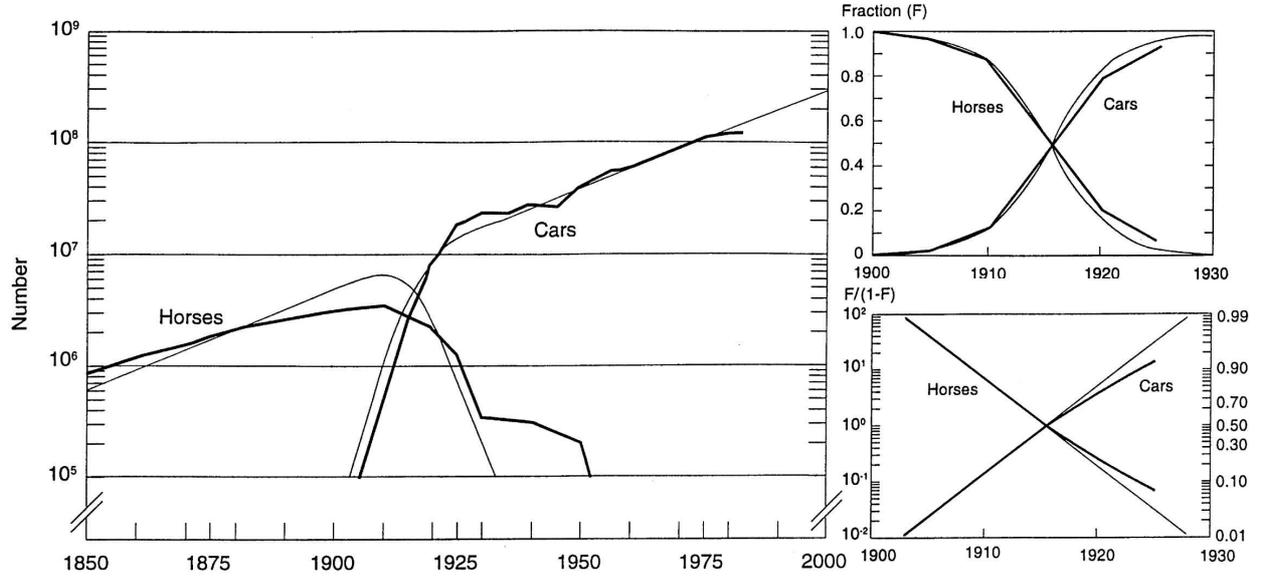}
		\end{center}
	\caption{Transition from horse-drawn carriages to petrol cars in the 1920s (data originally from \citealt{Nakicenovic1986}, graph taken from \citealt{Grubler1999}, reproduced with permission). (\emph{Left}) Raw data on a semi-log axis. (\emph{Top Right}) The data, when expressed as fractions of the total $F$, follows very closely logistic curves. (\emph{Bottom Right}) This demonstrated by a transformation of the data of the form $F/(1-F)$ on a semi-log axis, which produces nearly linear trends.}
	\label{fig:HorseCars}
\end{figure}

The parallel between technology and biology/ecology can be summarised as follows. Figure~\ref{fig:HorseCars} presents the iconic data from \cite{Nakicenovic1986} for the transition between horse-drawn carriages and petrol cars that occurred in the 1920s. In this data, a transition is observed superimposed onto an exponential growth in the number of vehicles. Through closer inspection, one observes that by dividing the numbers of horses and cars by the total number of transport units, functions reminiscent of logistic curves are observed that cross each other in around 1915 (using $S$ here for market \emph{Shares}):
\beq
S_1(t) = {1 \over 1 + \exp\big(\alpha_{12} (t-t_0)\big)}, \quad S_2(t) = 1- S_1(t) = {1 \over 1 + \exp\big(\alpha_{21} (t-t_0)\big)}.
\label{eq:logistic}
\eeq
This is shown to be an accurate assessment by displaying the fractional data as $S/(1-S)$ on semilog axes, generating linear trends, of which the time scaling parameters $\alpha$ are obtained from the slope:
\beq
\log \left({S_1 \over 1-S_1}\right) = \alpha_{12}(t-t_0), \quad \log \left({S_2 \over 1-S_2}\right) = \alpha_{21}(t-t_0), \quad \alpha_{12} = -\alpha_{21},
\eeq
Taking a time derivative of these expressions, one obtains a pair of differential equations fully describing the system:
\beq
{dS_1 \over dt} = \alpha_{12} S_1 \big(1 - S_1 \big) = \alpha_{12} S_1 S_2, \quad {dS_2 \over dt} = \alpha_{21} S_2 S_1.
\eeq
This example depicts the interaction occurring within a pair of technologies. \cite{Geels2005} criticises the analysis of \cite{Nakicenovic1986} by invoking the presence of two other important transport technologies that have interacted with and influenced the development of petrol vehicles but have not pervaded the market, namely electric trams and bicycles. Effectively, in most cases of technology competition, it is nearly impossible to exclude the existence of a third interacting component, and a fourth and so on,\footnote{The perverse effect of using quantities relative to the total is that this method can easily lead to overlooking other competing technologies that only hold small market shares.}

\beq
\left. \begin{array}{ll} \dot{S_1} = \alpha_{12} S_1 S_2 + \alpha_{13} S_1 S_3 + \alpha_{14} S_1 S_4+... \\ \dot{S_2} = \alpha_{21} S_2 S_1 + \alpha_{23} S_2 S_3 + \alpha_{24} S_2 S_4+...\\
\quad\quad\quad\quad\quad\quad\quad\quad\quad \vdots \\
\dot{S_n} = \alpha_{n1} S_n S_1 + \alpha_{n2} S_n S_2 + \alpha_{n3} S_n S_3+...
\end{array}\right\} \Rightarrow {dS_i \over dt} = \sum_{j=1}^n \alpha_{ij} S_i S_j,
\label{eq:replicator}
\eeq
generalising the theory to an arbitrary number $n$ of technologies interacting in the marketplace, with interaction time constants held in the antisymmetric matrix $\alpha_{ij}$. It corresponds to the so-called replicator dynamics used in evolutionary game theory \citep[imitation dynamics, ][p86]{Hofbauer1998} and evolutionary economics \citep{Safarzynska2010}, in binary interaction form. A more common version however is one of comparison of the \emph{fitness} of a candidate to the average fitness of the population,
\beq
{dS_{i} \over dt} =  S_i \sum_j \left( \mathcal{F}_i(S_j) - \overline{\mathcal{F}}(S_j) \right),
\label{eq:Creplicator}
\eeq
The replicator equation is mathematically equivalent to the Lotka-Volterra system of differential equations for the numbers of individuals in a set of competing species in an ecosystem when expressed in absolute numbers,
\beq
{dN_i \over dt} = r_i \left[ N_i - \sum_j {\alpha_{ij} N_i N_j \over N_{tot}} \right].
\label{eq:LV}
\eeq
Here, the first term $r_iN_i$ is the \emph{birth} of individuals with \emph{birth rates} $r_i$, generating an exponential growth component, and the second term, negative, expresses both the interference of a specie with itself, when resources become scarce and individuals begin to compete, or the interference across species\footnote{The relative signs of the elements in $\alpha_{ij}$ when permuting $i$ and $j$ determine the nature of the interaction, i.e. competition or predator-prey.} competing for the same resources. The new parameter $N_{tot}$ is the \emph{carrying capacity} of the ecosystem, the number of individuals that the system can accommodate. In the technology context, the carrying capacity corresponds to the total number of units of technology supplying the demand for a service, or societal function, following a \emph{demand led} economic assumption. In this analysis, however, the parameters $r_i$ and $\alpha_{ij}$ contain lots of information and thus need unpacking, which we proceed to do below.

\subsection{Combining technology demography and choice modelling}

This paper presents a model of technological change that explains the pattern given above, deriving from first principles a replicator dynamics equation for technologies \emph{at the unit level} from demography theory,\footnote{E.g. how long does a car survive for on roads? How many cars of a particular type can be produced in a year?} and provides meaning to its parameters ($\alpha_{ij}$) in terms of information that relates to technology and industry characteristics (e.g. life expectancy, rates of capital investments, etc). 
Several independent strands of demography exist, using either a continuous or a discrete form, \cite[all equivalent,][]{Keyfitz1977}, of which I shall choose the continuous form. Human demography in the continuous version corresponds to an age structured form of single specie population dynamics. It provides an in-depth view of the process of population evolution through age specific stochastic birth and death events, using probabilities of giving birth and of dying for age tranches covering a whole lifetime. This provides demographers with a finer accuracy for population projections than crude average birth and death rates. This is also partly equivalent to survival analysis as used in engineering to determine the statistics of failure of devices. A system of competing species can also be described with an age structure. I create here such a construction for technology dynamics, which, as I will show, explains the form of the technological change process due to its key property of \emph{self-correlation} in time.

In contrast to demography, however, the birth of technology obviously does not occur through pregnancy, although it is possible to define an equivalent birth function \citep[or maternity function, see][]{Kot2001}. Technology birth takes place in an industrial structure through the investment of financiers in production capital and labour, using for this the profits on sale of these same technologies. Sales are the process by which population expansions can take place: if sales increase, the production capital and labour can be expanded, but if sales decrease, the production capital and labour must eventually depreciate due to lack of investment. In order to explore this I thus proceed in section~\ref{sect:model} with describing mathematically the birth, death and the nature of competition in a market. 

The choice of technology however is a human process, and the human population is naturally diverse. To model choice by diverse agents, a standard theory exists which is commonly applied, discrete choice theory \citep[e.g. voter models, transport mode choice, see ][]{Ben-Akiva1985, McFadden1973}. This is also known as logit models, in which the choice probability commonly takes the form 
\beq
f_i = {e^{U_i} \over e^{U_i} + e^{U_j}}, (\text{Binary form}) \quad f_i = {e^{U_i} \over \sum_k e^{U_k}}, (\text{Multinomial form}),
\label{eq:logit}
\eeq
where $U_i$ is the so-called random (i.e. stochastic) utility associated to each choice. These models however, if applied as they are to technological change, fundamentally assume perfect information and technology access by diverse agents, and, without further dynamics, instantaneous diffusion. It however appears appropriate to connect discrete choice theory with the replicator equation of evolutionary game theory. As we show below, a duality exists between the binary interaction form of the Lotka-Volterra (eq.~\ref{eq:LV}) and the multi-technology form of the common replicator equation of evolutionary theory (eq.~\ref{eq:Creplicator}), and furthermore, the multinomial logit approximately emerges when transforming the Lotka-Volterra system in to this form of the replicator equation (section~\ref{eq:Derivation}).


\subsection{The multi-level perspective in transitions theory}

This work can be brought into the perspective of transitions theory, the main qualitative theoretical framework to describe transitions of socio-technical regimes. This paper however treats the problem from a multi-technology competition perspective, which can thus be brought into the transitions theory context. 

\begin{figure}[t]
		\begin{center}
			\includegraphics[width=.7\columnwidth]{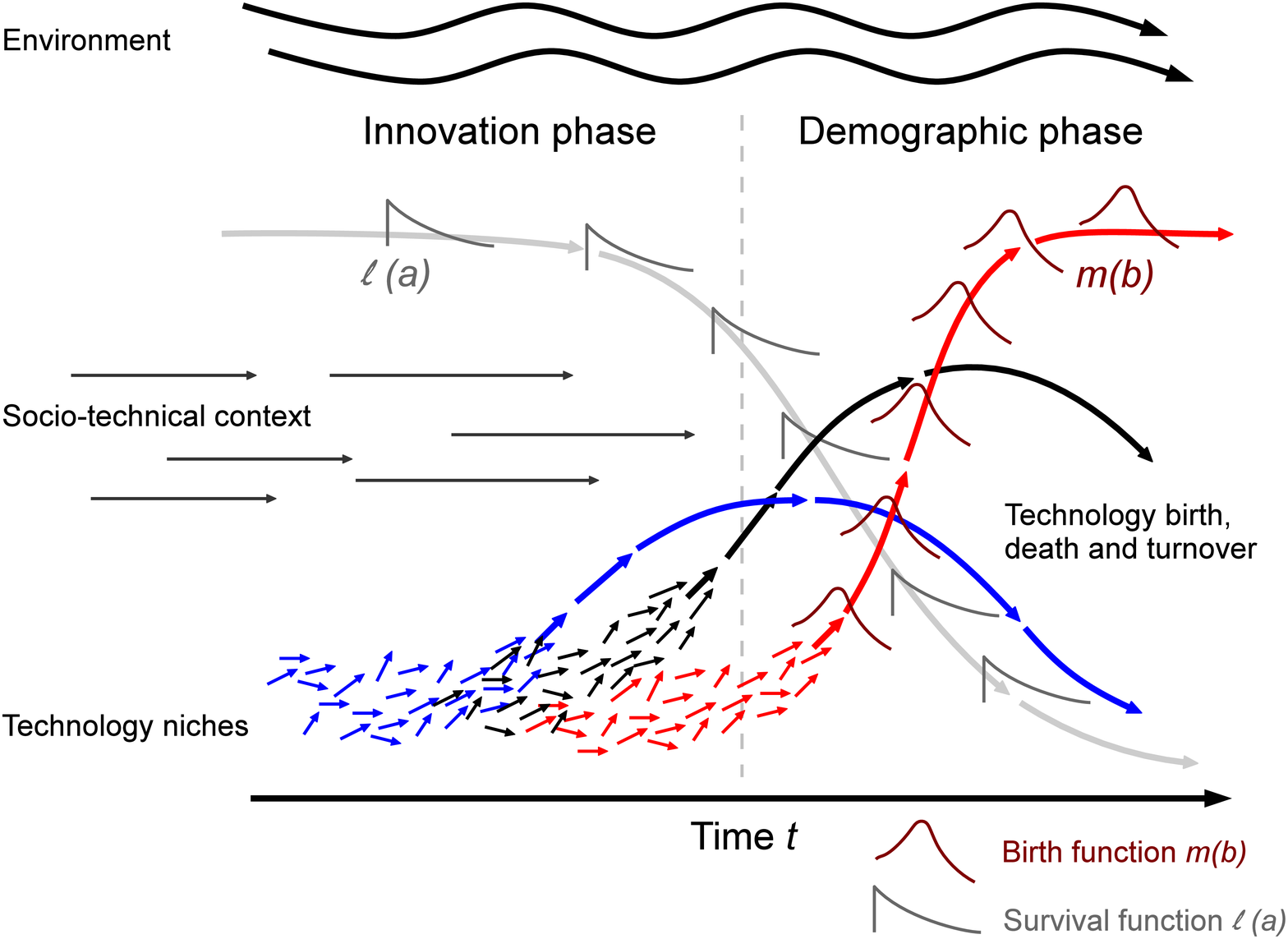}	
		\end{center}
	\caption{Illustration of the demographic phase of technology transitions, adapted from \cite{Geels2002}. Small arrows represent small incremental erratic innovations, during the innovation phase, which gradually gain coordination and momentum before diffusion takes place, entering the demographic phase. One technology is in decline (gray), disappearing at a maximum rate related to its survival function $\ell(a)$. One technology initially gains market shares at the expense of the one declining (blue), but is in time beaten in the race by another (black), which in turn is overtaken by yet another technology (red). The maximum growth rate is related to the birth function $m(b)$. The socio-technical context, consumer preferences and the environment generate selection mechanisms driving market share exchanges between technologies. }
	\label{fig:GeelsUpdate}
\end{figure}

Starting from the picture of \cite{Geels2002}, the process of technology transitions from a multi-technology competition perspective can be thought of as going through two different phases. This is depicted in figure~\ref{fig:GeelsUpdate}. New technologies originate from small, erratic, cumulative incremental innovations that gradually gain coordination as inventors and firms get to grips with understanding their own market and figuring out what is possible technically. This is shown with small randomly oriented arrows, with three colours indicating three innovations generating roughly the same service, or \emph{societal function}. Many trials and errors generate experience and learning that gradually determine the successful direction to take. Once this happens, better defined technologies in a particular socio-technical context begin to gain momentum of diffusion, and enter what I will call the \emph{demographic phase}. At this point, the growth rate is determined \emph{both} by: (1) agent choices (in terms of the respective advantages and flaws of competing technologies including the incumbent) within the socio-technical context and its evolution, (2) the timescales of birth and death, or technology turnover. In a situation of very clear and favourable consumer preferences and socio-technical evolution, the diffusion becomes limited by the birth rate of the new technology, and by the death rate of the old technology being replaced: the timescales of technological change which are the subject of this paper. 


The innovation phase is difficult to model in a forecasting context, as this would require knowing the unknown, innovations that have not yet been developed, inventions that have not yet been invented. Therefore, it is difficult to describe the emergence of new technologies beyond the qualitative picture by \cite{Geels2002}. However, the impacts of innovation on prices is simple to include using learning curves, if they are known. When technologies enter the demographic phase, they are well defined with a dominant design, and modelling their diffusion becomes straightforward, given a model of technology choices and knowledge of their survival properties, the birth and survival functions defined below, or equivalently the life expectancy and the rate of reinvestment into production capital and labour of all competing technologies. This diffusion is thus made uncertain in such a model in parts by the lack of possibility of emergence of disruptive alternatives. 



\section{Model components: birth, death and choice of technology \label{sect:model}}
\subsection{The death process}

\begin{figure}[t]
	\begin{minipage}[t]{.5\columnwidth}
		\begin{center}
			\includegraphics[width=.95\columnwidth]{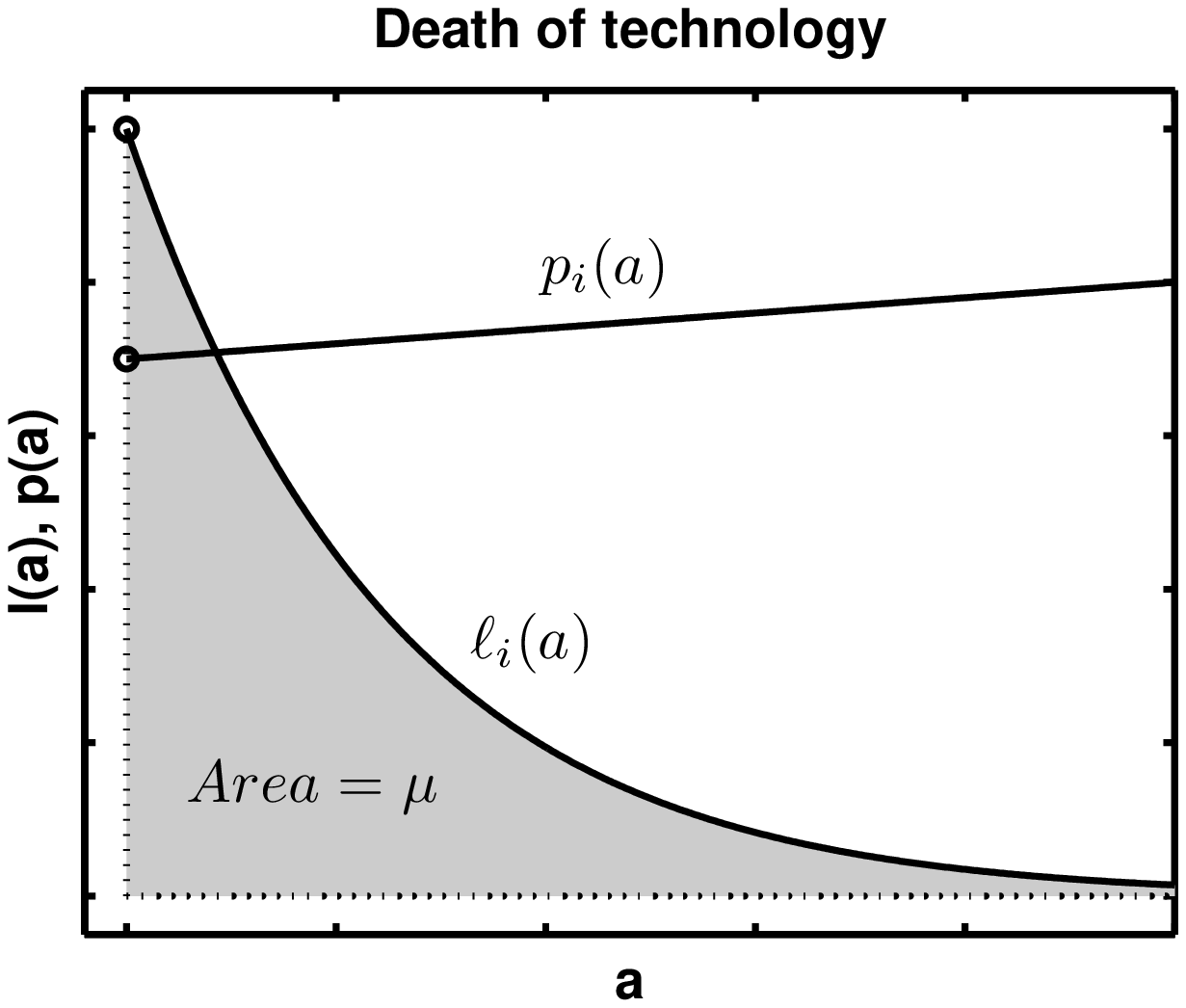}
		\end{center}
	\end{minipage}
	\hfill
	\begin{minipage}[t]{.5\columnwidth}
		\begin{center}
			\includegraphics[width=.95\columnwidth]{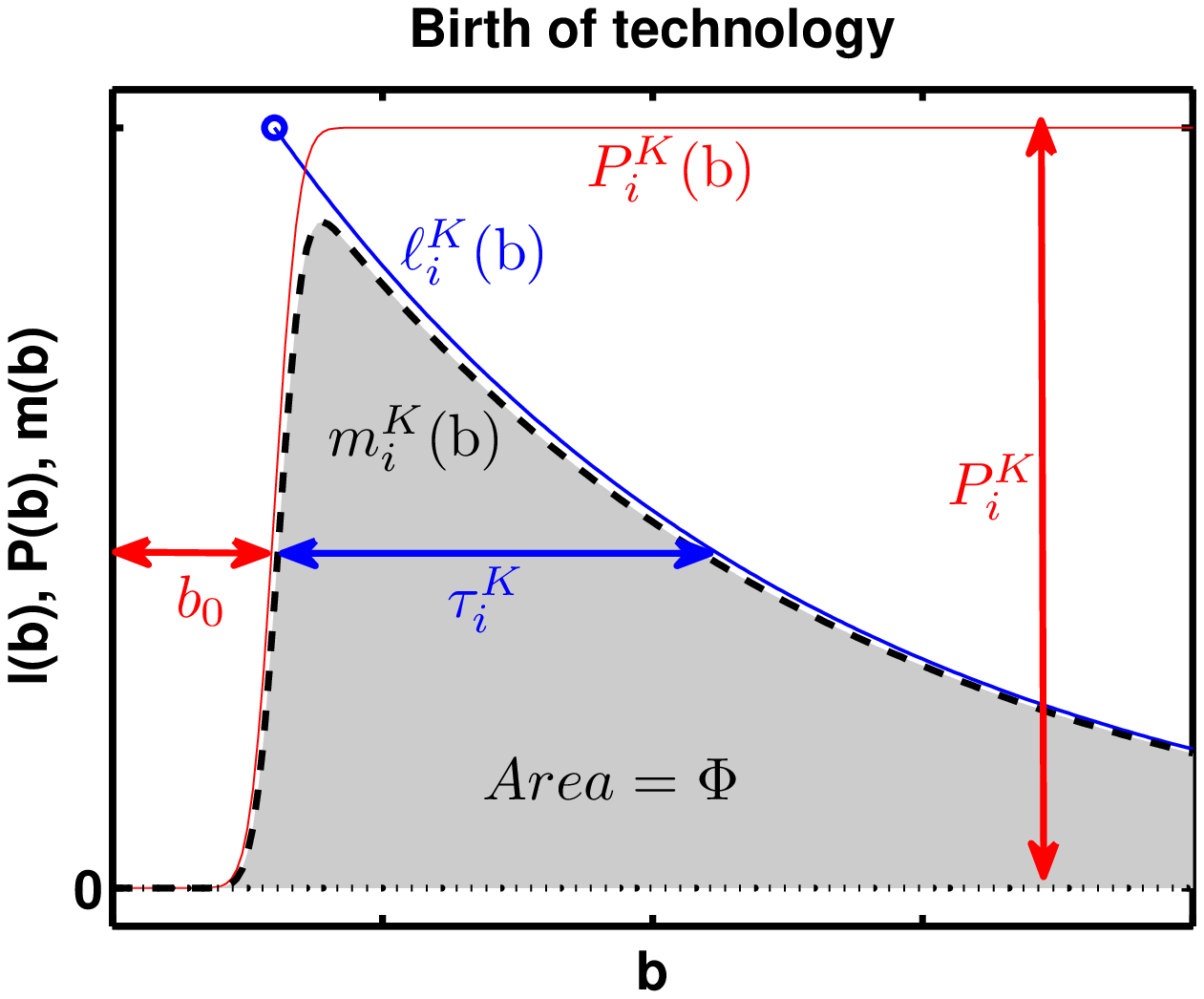}
		\end{center}
	\end{minipage}	
	\caption{Schema of the typical technology survival function $\ell_i(a)$ (\emph{left}) and birth function $m(b)$ (\emph{right}), which respectively stem from the instantaneous force of death $p(a)$ and a product of the production capital survival function $\ell_i^K(b)$ and the productivity $P_i^K(b)$. The area under $\ell_i(a)$ corresponds to the life expectancy $\tau_i$, while the area under $m_i(b)$ is the total expected production from one unit of capital during its lifetime, $\Phi_i$.}
	\label{fig:SurvivalFunction}
\end{figure}

The death of technology at the unit level can occur in different ways with different probabilities. For example, in the transport sector, vehicles can be retired due to fatal accidents, failures, or by economic decisions of owners due to increasing costs of maintenance with age.\footnote{The existence of \emph{sunk costs}, investments in exchange for which technology is expected to operate for a certain amount of time, imply the existence of a non-zero life expectancy. This is particularly true if money is borrowed for the purchase of a unit and repaid during its operating lifetime.} These processes have different probabilities of occurring as functions of vehicle age. For a technology of brand-model $i$, taking the probability of destruction at age $a$ as $p_i(a)\Delta a$, and the number $n_i(a,t')\Delta t'$ of technology units produced between year $t'$ and $t'+\Delta t'$ (or age interval $\Delta a$),\footnote{E.g. the number of 2003 Citroen C3 currently 11 years old.} the change in this age distribution of technology units $\Delta n(a,t)$ at time $t$ during an ageing interval $\Delta a$ due to destructions is
\beq
\Delta n_i(a,t')\Delta t' = -p_i(a) n_i(a,t')\Delta t' \Delta a.
\eeq
In the continuous limit ($\Delta a \rightarrow 0$), solving this for $a$ yields
\beq
n_i(a,t') \Delta t'= n_i(0,t') \ell_i(a) \Delta t', \quad \ell_i(a) = \exp\left(- \int_0^a p_i(a') da' \right).
\eeq
$\ell_i(a)$ is the common demographic \emph{survival function}, while $p_i(a)$ is the instantaneous \emph{force of death} \citep[see for instance][]{Keyfitz1977}. This is depicted in fig.~\ref{fig:SurvivalFunction} (left panel). When used in relation to people, survival functions are derived from life tables where individuals are traced during their lifetime from birth until death, which, when applied to technology, is called survival analysis in engineering. The various processes of technology death can be associated to components in $\ell_i(a)$. Accidents normally have a constant force of death, and therefore give $\ell_i(a)$ a simple exponential form. Meanwhile, scrapping due to failures tend to occur later during technology life, with increasing values of $p(a)$. Thus $\ell(a)$ can be written as
\beq
p(a) = {1 \over \tau_1} + {a \over \tau_2^2} + {a^2 \over \tau_3^3} + ... , \quad \ell(a) = \exp\left( - {a \over \tau_1} - {a^2 \over 2 \tau^2_2} - {a^3 \over 3 \tau^3_3} - ... \right),
\eeq
each term corresponding to different destruction processes with different timescales $\tau_n$. If accidents dominate the destruction process, then $\ell_i(a)$ should take predominantly an exponential form, while if the probability of failures dominates and increases approximately linearly with age, $\ell_i(a)$ takes the form of a gaussian, and so on. The survival of transport vehicles in the USA was shown to follow approximately a mixture of $\tau_1$ and $\tau_2$ processes \citep[][and references therein]{ORNL2012}. While $\ell_i(a)$ expresses the probability of a technology unit to remain in use until age $a$, the negative of its derivative expresses the probability of destruction at age $a$. The life expectancy $\tau_i$ is defined as 
\beq
\tau_i = -\int_0^\infty a {d\ell_i(a) \over da}da = \int_0^\infty \ell_i(a)da,
\label{eq:lifeexp}
\eeq
where the last expression above is obtained from the previous by integration by parts. In the simple case of death dominated by accidents, $\tau_i = \tau_1$ and units of a particular age tranche decrease in numbers exponentially at a rate equal to the life expectancy.

Every year $t = t'+a$, a certain number of deaths $d_i(t)$ occur, technology units that are scrapped in some way or another, while a number $\xi_i(t)$ of new units are sold, both changing the total number of units in use,
\beq
{\Delta N_i \over \Delta t} = \xi_i(t) - d_i(t).
\eeq
While deaths decrease the number of units of all ages, sales generate units of age zero. The gradual decease in numbers with ageing is
\beq
{d n_i(a,t') \over d t}\Delta t' = n_i(0,t') {d\ell_i(a) \over da}\Delta t', \quad n_i(0,t') = \xi_i(t'),
\eeq
where for each age tranche between $a$ and $\Delta a$ (or production year between $t'$ and $t' + \Delta t'$), the number of deaths depend on the probability of destruction times the number of units of that age remaining, which decreases every year. The number of units in each age tranche originates from sales that happened $a$ years ago (in year $t'$). Thus, in the continuous limit $\Delta t' \rightarrow 0$, while the total number of units at time $t$ depends on the number of units sold in the past that still remain at time $t$,
\beq
N_i(t) = \int_{-\infty}^t \xi_i(t') \ell_i(a) dt' = \int_0^\infty \xi_i(t-a) \ell_i(a) da,
\label{eq:living}
\eeq
the reduction in the number of units at year $t$ due to deaths is the sum of the number of units that remain in each age tranche and their probability of being destroyed precisely in year $t$,
\beq
 d_i(t) = - \int_0^\infty \xi_i(t-a) {d \ell_i(a) \over da} da.
 \label{eq:death}
\eeq

These expressions are the first and second \emph{convolutions} encountered in this theory, of past sales with the survival function and the probability of death. If sales $\xi_i(t)$ are related in any way to the existing number of units, this produces a \emph{self-correlation} of the number of units with itself in past years. I will show later that this is effectively the case, which restricts how fast the total number of units can change in the system.\footnote{Note that in contrast to the birth function that I shall define further, $\ell_i(a)$ must be a strictly decreasing function of age otherwise dead units would come back to life.}

\subsection{The birth process \label{sect:birth}}

The number of units of technology that can be built in a time span depends on the production capital and labour available at that time. However, production capital also wears out as it ages (when it is not repaired) and has a finite lifetime, and therefore its own survival function, which we denote $\ell_i^K(b)$, with age variable $b$. The production capital $K(b,t')$, installed at time $t'$ of age $b$, will begin production after a certain delay of installation $b_0$, and therefore its age dependent productivity function, $P^K_i(b)$, is zero at $b=0$ (see fig.~\ref{fig:SurvivalFunction}). The production decreases statistically with age however, since production capital gradually break down with ageing following $\ell_i^K(b\rightarrow \infty) = 0$.\footnote{We consider here repairs as investments in new production capital, in order to correctly keep track of the amount of depreciation.} The number of units of technology produced per year by these production units of age $b$ at time $t$ during an interval of capital ageing is therefore $K(b,t) \ell_i^K(b) P_i^K(b) \Delta b$.

Investment in new units of production capital is carried out using part of the income from the sale of produced technology units (we assume, of technology of the exact same type, we do not mix funds across different industries). Taking $R_i$ as the fraction of re-investment of profits into production capital,\footnote{$R_i$ is in units of production capital purchased per unit of technology sold.} the amount of capital of age $b$ at time $u$ scales with sales that occurred $b$ years earlier, i.e. $K(b,t') = R_i \, \xi_i(t') = R_i \, \xi_i(t-b)$. The total production capacity $\delta N_i(t)$ of all vintages can be calculated from the amount of capital that was built with funds from sales in all previous years. Defining the technology \emph{birth} function $m_i(b) = P_i^K(b)\ell_i^K(b)$, this is
\beq
\delta N_i(t) = R_i \int_0^\infty \xi_i(t-b)m_i(b) db,
\label{eq:ProdCap}
\eeq
where the difference between the production capacity and actual sales depends on the presence of competitors and consumer choices.\footnote{Whether the capital is fully used or whether there is spare capacity.} This is the third convolution of this theory, which generates, if the number of production units is related to sales, another autocorrelation in the number of units. 

As opposed to $\ell_i(a)$, $m_i(b)$ it is not a strictly decreasing function, but it increases initially, as production begins some time after construction, before decreasing in later years when old production lines get decommissioned. It must be an integrable function, the area under which $\Phi_i = \int_0^\infty m_i(b) db$ converges. As we show in section~\ref{sect:Approx}, in order to have an increasing production capacity, we must have $R\Phi_i > 1$.

The fastest possible rate of growth of sales can be calculated by hypothesising a fictitious situation without competition where households are able to consume any level of production, therefore with indefinite growth (a fictitious purely supply-led market\footnote{It almost never happens that a rate of production growth determined solely by the supply side persists for a long time. For example in the transport sector, if sales in developed nations were to increase faster than the population, this would mean that households eventually own 3-4-5 cars and so on, rather unlikely. In this case, the rate of growth of sales is limited by the rate of growth of the demand, not the rate at which production could hypothetically be scaled up given its profitability. Supply-led growth however could arise in special circumstances such as in wartime policies of rapid up-scaling.}). The production capital is under full employment and the total amount of production of technology units is 
\beq
\xi_i(t) =  R_i \int_0^\infty \xi_i(t-b)\ell_i^K(b) P_i^K(b) db.
\eeq
Thus, in this monopolistic case sales that occur in the present are completely determined by sales in the past, where the income on past sales were used to expand the production capacity, which enables more production and more sales, and thus more expansion and so on. This is identical to renewal equations in demography where birth rates in the present depend on what birth rates have been in the past \citep{Keyfitz1977, Kot2001}, also called Lotka's integral equation \citep{Lotka1911}. This leads to exponential solutions with possible oscillatory components \citep{Keyfitz1967} for both $\xi_i(t)$ and $K(t)$, therefore indefinitely increasing sales and capital. Obviously, such a monopoly could never be maintained indefinitely and sales must presumably fall short of production at some point in time where a competitor interferes with the market with a more successful product. However, in a situation where an innovation were to take such a path free of competitors, it would follow the fastest possible rate $t_i^{-1}$ of growth determined in \ref{sect:AppA}, equal to either:
\begin{enumerate}
\item $R_i P_i$, the rate of investment in the case where the production capital lifetime is long-lived and only a short or no delay takes place in its construction,
\item $b_0$, the time delay, in case a long delay takes place in the construction of the production capital.
\end{enumerate}

\subsection{The choice process}

\begin{figure}[t]
		\begin{center}
			\includegraphics[width=.7\columnwidth]{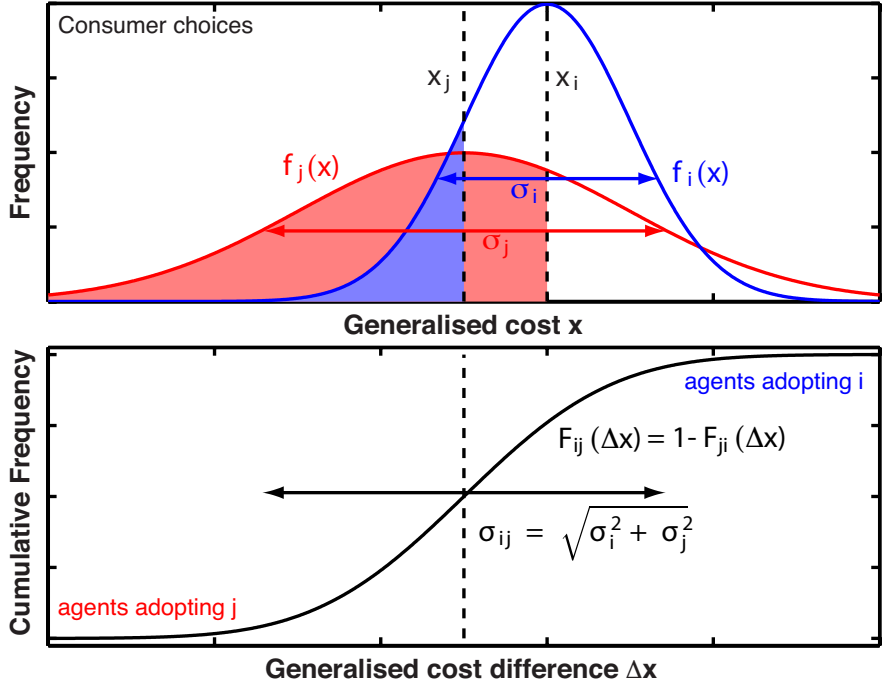}
		\end{center}
	\caption{Illustration of the process of decision-making under diversity of agents between two technologies. The blue curve represents the distribution of perceived generalised costs for one technology, and the red curve that of the other. In the left panel, if diversity is very low, choices can flip very abruptly as average costs cross. This corresponds to the representative agent case. In the right panel, introducing significant diversity makes choices distributed and choices change very gradually as costs cross.}
	\label{fig:Choices}
\end{figure}

As we know from diffusion theory \citep{Rogers2010}, the diversity of agents is linked to diffusion rates. I thus create a model of decision-making in the context of diverse agents. For a model of technology diffusion, we require an aggregate representation of decision-making when agents are diverse, and costs have variations. Diversity stems from different perceptions from agents when they take a decision, which may originate from a large set of particular preferences and constraints that is impossible to enumerate in a model. We require this diversity to be summarised by distributions. For this, I assume that choice is made on the basis of a single quantity, a generalised cost $x$ (figure~\ref{fig:Choices} top), evaluated by agents for each option they see as available to them, and this value must feature a quantification of all possible aspects that weigh in the decision-making balance. 

To clarify, I postulate here that \emph{distributions of perceived costs correspond to distributions of observed costs}. I justify it as follows: agents, I assume, when considering investing in a unit of technology (e.g. a car), most likely choose something they have seen chosen before, perhaps by someone they know, such that they were able to gather information.\footnote{i.e. they most likely do not choose something they know nothing of, and they gather reliable information predominantly through observations of their peers} This may be due to their belonging to a particular social group and social class, and they are most likely to choose amongst what their peers have previously chosen, which itself is a subset of what the whole market has to offer (e.g. poor rural households perhaps purchase different types of vehicles to rich suburban families, which itself is different than single middle-class persons, i.e. their peers are a subset of the population and their observations are a subset of all observations).  This is a key part of what is known as the anthropology of consumption \citep{Douglas1979}. Thus I assume \emph{restricted technology/information access}, in other words, agents do not choose what they do not know, and they do not know, or care for, all options technically available. Choices of particular social groups endure through peer observation and visual influence, which has been demonstrated empirically for for example with vehicle purchases in the USA \citep{Mcshane2012}. 

The frequency of events of observations of a particular technology model (by consumers shopping), sample of an ensemble of such events, corresponds to the frequency of recent sales of that model (purchases by their peers). It follows that the probability of choosing a particular model is most likely proportional to this frequency of observation, and thus these preference distributions, associated to circumstances, constraints and social group origin of consumers, difficult to enumerate and unknown to the modeller, are relatively stable. These combined frequencies form a generalised cost distributions of sales. Therefore in this perspective, \emph{the generalised cost distribution of recent sales is a representation of the diversity of choices}. I shall go further and claim that we can use the measured diversity of sales and interpret it in terms of the diversity of agents. In such a perspective, technology sales by type and model reinforce technology sales of those types and models, consistent with the work of \cite{Mcshane2012}. This is a situation of \emph{increasing returns to adoption}, discussed by \cite{Arthur1989}, which, combined with diversity, he demonstrates leads to path dependence and several equilibrium points. 

This approach to modelling choice is not new, termed discrete choice theory \citep{Ben-Akiva1985, McFadden1973}, where the generalised cost being minimised (figure~\ref{fig:Choices}) is equivalent to the random utility being maximised in the more classical version of the theory. In binary logit form, the description of decision-making is represented here using pairwise comparisons of cost distributions (figure~\ref{fig:Choices}, top). When faced with equivalent technology choices, the fraction of investors or consumers choosing technology $i$ over $j$ can be approximated by counting the number of instances out of the total where the generalised cost of technology $i$ falls below every possible other value of the generalised cost of technology $j$, and vice-versa. Following standard discrete choice theory if the distributions are of the Gumbel type (extreme value distributions), with standard deviations (diversity) $\sigma_i$ and $\sigma_j$, then the frequency where technology $i$ is less costly than technology $j$ follows a logistic distribution (the binary logit in eq.~\ref{eq:logit}), of width parameter $\sigma_{ij} = \sqrt{\sigma_i^2 + \sigma_j^2}$. The pairwise comparison generates a choice likelihood of adopting $i$ over $j$ that I denote $F_{ij}$, and a likelihood of adopting $j$ over $i$ $F_{ji} = 1 - F_{ij}$. This is derived exactly and explained further in \ref{sect:AppB}.

Finally, one comment may be added concerning innovation. Technologies are not static but change as production methods improve, and these improvements occur through re-investment and production. Innovation in the firm leads to learning-by-doing cost reductions, which influences the choices of agents \cite[e.g.][]{McDonald2001, Weiss2012}. In particular, the costs of new technologies typically change faster than those of the incumbent. Learning curves, of the usual form $C_i(t) \propto (\int \xi_i(t) dt)^{-b}$ (cumulated sales) can easily be included in the decision-making process described here.


\section{An age structured model of technology competition \label{sect:model2}}
\subsection{Deaths replaced by births}

I now build a model of technology competition and substitution. Choices of consumers or investors are taken to mean what choices would be made if all options were not equally available or known to all agents. This is defined in terms of preferences in the comparison of each possible pairs of technologies $F_{ij}$. Given that despite the first choice of consumers or investors, those may not necessarily be available in every individual situations due to the amount of existing industrial capacity to produce them, consumers or investors may have to content themselves with their second or third choice. This is important here since we are dealing with technology diffusion, and that the diffusion process involves a gradually changing availability of new technologies.

Considering that units of any age are replaced by new ones when they come to the end of their life and are scrapped,\footnote{i.e. accidents, breakdowns or economic scrapping decisions, following the survival function. The nature of ownership of these technology units, and whether they change ownership, is not relevant, which enables to make abstraction of second-hand markets.} following this approach, I evaluate the number of units removed from one arbitrary technology category $j$ and added into another category $i$. For this, I start with the total number of deaths to be replaced in all vehicle categories and ages, and find how many of those belong to category $j$. Out of those destructions in $j$, I evaluate those that were chosen by consumers to be replaced by technology $i$, according to $F_{ij}$. 

Of these, only a fraction can be produced. The production capacity of a particular technology may not necessarily be able to supply the demand in every one of these situations, were the consumers to all simultaneously choose this technology. Therefore, in a certain number of these situations, the option will simply not be available, and consumers will have to choose between the remaining options despite their best preference. The fraction of instances where this choice will be available with respect to the total number of choices being made corresponds to the fraction of production capacity of this technology with respect to the total production capacity. 

This can be understood through an analogy involving an ensemble of shops with a number of competing products on their shelves, and agent only go to their local shop, and thus each see a different set of options. Given the production capacity of each product's respective industry, most shops will not be able to stock units of all competing products. The relative frequency of shops stocking particular technology models corresponds to the relative production capacity for those models.  When customers have equal preferences for all products, the relative probability of the average customer choosing particular products corresponds to the \emph{average} composition of the product choice in the ensemble of shops, which itself corresponds to relative production capacity of each product with respect to the total. Thus the fraction of units of technology $j$, chosen to be replaced by technology $i$, that can actually be replaced by units of $i$ corresponds to the fraction of the total production capacity that produces technology~$i$. 

I define a flow of units from categories $j$ to $i$ as follows:
\beq
\Delta N_{j\rightarrow i} = \left[ \begin{array}{ll} \text{\small Fraction of} \\ \text{\small prod. capital} \\ \text{\small belonging to \emph{i}} \end{array}\right]_i
\left[ \begin{array}{ll} \text{\small Consumer} \\ \text{\small preferences} \end{array}\right]_{ij}
\left[ \begin{array}{ll} \text{\small Fraction of} \\ \text{\small deaths} \\ \text{\small belonging to \emph{j}} \end{array}\right]_j
\left[ \begin{array}{ll} \text{\small Number of} \\ \text{\small deaths} \end{array}\right]_{tot}
\label{eq:Words}
\eeq

The net flow from both directions between $i$ and $j$ (some agents make opposite choices), $\Delta N_{ij}$, and the sum of all changes for any technology $i$, are:
\beq
\Delta N_{ij} = \Delta N_{j \rightarrow i} - \Delta N_{i \rightarrow j}, \quad \Delta N_i = \sum_j \Delta N_{ij}.
\label{eq:exchange}
\eeq

\subsection{The age structured model}

Eq.~\ref{eq:Words} can be written in terms of the production capacity $\delta N_i(t)$ and deaths $d_j(t)$, as defined above:
\beq
\Delta N_{j \rightarrow i} = \left( {\delta N_i(t) \over \sum_k \delta N_k(t)}\right) F_{ij} \left( { d_j(t) \over \sum_k d_k(t) }\right) \left( \sum_k d_k(t) \right) \Delta t.
\eeq
The production capacities and death numbers at time $t$ can be replaced by convolutions of past sales (eqns~\ref{eq:death} and \ref{eq:ProdCap}):
\beq
\Delta N_{j \rightarrow i} = \left( {R_i \int_0^\infty \xi_i(t-b) m_i(b) db \over \sum_k R_k \int_0^\infty \xi_k(t-b) m_k(b) db}\right) F_{ij} \left( { \int_0^\infty \xi_j(t-a) {d \ell_j(a) \over da} da \over \sum_k \int_0^\infty \xi_k(t-a) {d \ell_k(a) \over da} da }\right) \nn
\eeq
\beq
 \left( \sum_k \int_0^\infty \xi_k(t-a) {d \ell_k(a) \over da} da \right) \Delta t.
\label{eq:notWords}
\eeq
Note the symmetry between the production side and the destruction side of this equation. There is, effectively, a high similarity between both processes. The difference however is fundamental: while $\ell(a)$ is a strictly decreasing normalised function of age and smaller than 1, generating destruction only, $R_i m(b)$ increases, and its integral is greater than 1, generating production. However, in order not to have an indefinitely increasing production capacity, $m(b)$ also decreases again at high values of $b$, maintaining the function integrable\footnote{The production capital produces a finite amount of goods in its lifetime.}, generating decreases in the production capacity when sales decrease, reflecting the gradual depreciation and wearing out of the production capital if no funds are available to replace them.

Eq.~\ref{eq:exchange} with eq.~\ref{eq:notWords} provide an expression for exchanges of units between categories (the exchange term). However, the total number, the carrying capacity, could also be changing, requiring either units that are brought in that do not replace deaths, or deaths that are not replaced. In the more common case of a total population $N_{tot}$ increasing, this is met by technology production following the relative production capacity:
\beq
\Delta N_i^\uparrow = \left( {R_i \int_0^\infty \xi_i(t-b) m_i(b) db \over \sum_k R_k \int_0^\infty \xi_k(t-b) m_k(b) db}\right)  \left({\Delta N_{tot} \over \Delta t}\right) \Delta t,
\eeq
where choices need not be involved.\footnote{Adding here a factor $F_{ij}$ can be done but is secondary: even if new units are not chosen exchanges can occur through the exchange term.} Meanwhile in the second less common case, the decrease in $N_{tot}$ is met by the relative rate of deaths,
\beq
\Delta N_i^\downarrow = \left( { \int_0^\infty \xi_i(t-a) {d \ell_i(a) \over da} da \over \sum_k \int_0^\infty \xi_k(t-a) {d \ell_k(a) \over da} da }\right) \left({\Delta N_{tot} \over \Delta t}\right) \Delta t.
\eeq
Assembling these expressions together, one obtains an expression too large to write here, summarised by 
\beq
\Delta N_i = \sum_j \Delta N_{ij} + \Delta N_i^\uparrow \quad \text{or}\quad \Delta N_i = \sum_j \Delta N_{ij} + \Delta N_i^\downarrow.
\label{eq:total}
\eeq

When terms are replaced in eq \ref{eq:total}, the resulting large expression corresponds to the demographic model of technology expressed in terms of the full sales history. This is the most general model of technology competition that can be derived from deterministic demography theory.\footnote{Thus improvements could be made using stochastic population growth theory, where for instance the probability of extinction at low population numbers would be better represented.}

This model can also be expressed uniquely in terms of sales, where $\xi_i(t) = \sum_j \Delta N_{j \rightarrow i} + \Delta N_i^\uparrow$:
\beq
\xi_i(t) = \sum_j \left( {R_i \int_0^\infty \xi_i(t-b) m_i(b) db \over \sum_k R_k \int_0^\infty \xi_k(t-b) m_k(b) db}\right)  F_{ij} \left( \int_0^\infty \xi_j(t-a) {d \ell_j(a) \over da} da \right)\nn
\eeq
\beq
 + \left( {R_i \int_0^\infty \xi_i(t-b) m_i(b) db \over \sum_k R_k \int_0^\infty \xi_k(t-b) m_k(b) db}\right) \left({\Delta N_{tot} \over \Delta t}\right)
\label{eq:Renewal}
\eeq
This fully recurrent population growth equation expresses how sales in the \emph{present} are constrained by \emph{sales in the past} within and between categories, through convolutions, generating self and cross-correlations of the sales. Since sales are autocorrelated in time, and that the addition of units corresponds to sales and removals to deaths, it implies that the absolute numbers of units are self and cross-correlated in time as well. Therefore, changes in the numbers of units cannot happen faster than is allowed by these correlations, which as we demonstrate next, are given by the length in time of the functions $\ell_i(a)$ and $m_i(b)$. 

Going any further requires evaluating all the convolutions, which would involve full knowledge of sales $\xi_i(t)$ in addition to survival functions $\ell_i(a)$ and birth functions $m_i(b)$. This equation can however be simplified enormously with the two following approximations.

\subsection{Simplification of the model with key approximations\label{sect:Approx}}

\begin{figure}[t]
		\begin{center}
			\includegraphics[width=0.5\columnwidth]{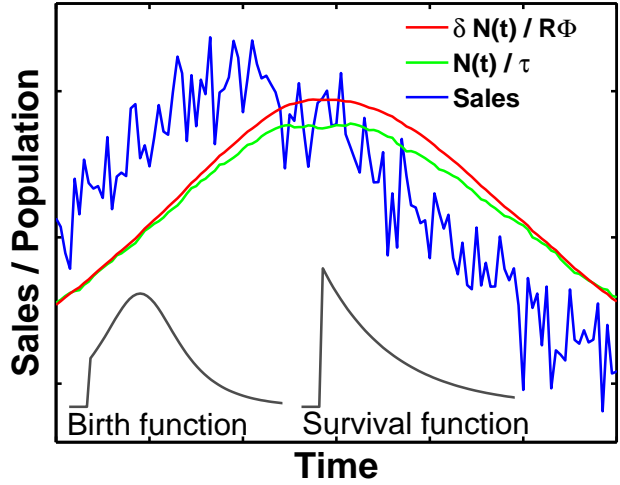}	
		\end{center}
	\caption{Computational experiment illustrating the convolution sales with the birth and death functions, justifying both approximations of this section.}
	\label{fig:Convolutions}
\end{figure}

Eq.~\ref{eq:total} in its full form, or alternatively eq.~\ref{eq:Renewal}, appear rather complicated, unconstrained and un-instructive. However, since they are recurrent, these equations are more constrained in terms of their possible solutions than they seem. Eq.~\ref{eq:total} expresses technological change between technology categories in terms of respective sales of those technologies. These sales are convolved with the functions $m(b)$ and $d\ell(a) / da$. It is well known in signal processing theory that convolutions of time series with functions of bounded kernels (fig.~\ref{fig:Convolutions}) yield slightly modified time series that are \emph{smoothed} with respect to the original, where high frequency changes have been damped.\footnote{i.e. a `low-pass' filter.} The `cutoff' value at which frequencies are suppressed, the sharpness limit, corresponds to the \emph{width} in time of the kernel.\footnote{In this case both $m(b)$ and $d\ell(a)/da$; the wider the kernel, the lower the frequency cutoff and the more smoothing occurs.} This is also the \emph{correlation length} of the smoothed function. For symmetrical normalised kernels of similar widths but different shapes, the convolution of a function leads to very similar results since a similar frequency cutoff occurs, and the same amount of damping occurs. If a kernel is not normalised, it either amplifies the time series (its integral is greater than one) or damps it (its integral less than one). If both kernels are not normalised but of similar widths, the convolutions will yield results which are close to multiples of each other, \emph{with proportionality factor the relative area under the kernels}.\footnote{This results from the convolution theorem.} Finally, if the kernels are not symmetrical functions, as is the case here, a time offset may appear between the two convolved outputs. This is demonstrated with a computational example shown in figure~\ref{fig:Convolutions}, where a hypothetical noisy time series (sales) was convolved with hypothetical birth and death functions. The result is almost independent of the shape of these functions, except for a proportionality factor, the relative area under the two kernel functions (the time offset results from the asymmetry in time of the kernels). In our case here with the birth and survival functions as kernels, this relative factor is $R_i\Phi_i/\tau_i$.

\vspace{8pt}
\subsubsection*{Approximation 1: the shape of the death function}

The first kernel, $d\ell_i(a)/da$, is normalised by definition (expressing an eventual but certain death), while the life expectancy is defined by eq.~\ref{eq:lifeexp}:
\beq
\int_0^\infty -{d\ell_j(a)\over da} da = 1, \quad \int_0^\infty \ell_j(a) da = \tau_i.
\eeq
Since the shape of the kernel does not matter much for the convolutions as long as its width in time is maintained, for our purpose we can approximate that,
\beq
 -{d\ell_j(a)\over da} \simeq {\ell_i(a) \over \tau_i}.
\eeq
This means that from eq.~\ref{eq:living}, which relates numbers to sales through the survival function, the convolution for deaths in eq.~\ref{eq:death} becomes
\beq
\int_0^\infty \xi_j(t-a) {d\ell_j(a) \over da}da \simeq {1\over \tau_i} \int_0^\infty \xi_j(t-a) \ell_j(a) da = {N_j(t) \over \tau_i},
\eeq

\vspace{8pt}
\subsubsection*{Approximation 2: similarity between the birth and death functions}

The second kernel, the birth function $m_j(b)$, has the following property,
\beq
R_i \int_0^\infty m_i(b) db = R_i\Phi_i > 1,
\eeq
which reflects the growth of the production capacity through reinvestment.\footnote{The production of goods using existing capital generates more wealth than just what is required to maintain itself.} 
In a case where the width of the second kernel, the birth function $m_i(b)$, is similar to the width of the survival function $\ell_i(a)$ (or alternatively the death function $-{d\ell_i(a) \over da}$), the convolution of sales by ${d\ell_i(a) \over da}$ or $m_i(b)$ will not be very different, but rather approximately proportional. Conversely, if the widths are very different, they \emph{cannot} in any way be proportional or even similar. The width of the birth function is related to the survival function of the capital and labour used for production, the production lines, which may have, in some situations, a similar time scale. Assuming that this is so (i.e. $\tau_i \simeq \tau_i^K$), and since $-{d\ell_i(a) \over da}$ is normalised, then the convolutions with $m_i(b)$ and $-{d\ell_i(a) \over da}$ are approximately proportional, and the proportionality factor is $R_i \Phi_i$:
\beq
R_i \int_0^\infty \xi_i(t-b) m_i(b) db \simeq - R_i \Phi_i \int_0^\infty \xi_i(t-a) {\ell_i(a) \over \tau_i} da = {N_i(t) \over t_i} \nn
\eeq
\beq
\text{where } t_i = {\tau_i \over R_i\Phi_i} \simeq {1 \over R_i P_i},
\label{eq:Approx2}
\eeq
$t_i$ is the newly defined fastest possible timescale of growth of the production capacity before considering investor choices, consistent with the result of \ref{sect:AppA} (section~\ref{sect:conditions}).

Thus with these approximations, I can replace the convolutions in eqns. \ref{eq:notWords} and \ref{eq:total} by $N_j/t_j$ and $N_i/\tau_i$, which considerably simplifies the system of coupled equations. 

\subsection{Derivation of the replicator and Lotka-Volterra equations \label{eq:Derivation}}

Substituting in eqns. \ref{eq:notWords} each convolution by its associated approximation, I obtain
\beq
\Delta N_{j \rightarrow i} = \left( { {N_i(t) \over t_i} \over \sum_k {N_k(t) \over t_k} }\right) F_{ij} \left( { {N_j(t) \over \tau_j} \over \sum_l {N_l(t) \over \tau_l} }\right) \left( \sum_m {N_m(t) \over \tau_m} \right) \Delta t,
\label{eq:evenlesswords}
\eeq
Defining the population weighted average frequencies $\overline{t}^{-1}$ and $\overline{\tau}^{-1}$,
\beq
{1 \over \overline{t}} = {1 \over N_{tot}} \sum_k {N_k(t) \over t_k} \quad \text{and}\quad  {1 \over \overline{\tau}} = {1 \over N_{tot}} \sum_l {N_l(t) \over \tau_l},
 \eeq
 the flow becomes
\beq
\Delta N_{j \rightarrow i} = \left( {\overline{t} \over t_i }{ N_i(t) \over N_{tot} }\right) F_{ij} \left( {\overline{\tau} \over \tau_j } { N_j(t) \over N_{tot}  }\right) \left( {N_{tot}\over \overline{\tau}} \right) \Delta t,
\eeq
while the term concerning increases in carrying capacity $N_{tot}$ becomes
\beq
\Delta N_i^\uparrow = \left( {\overline{t} \over t_j } { N_i(t) \over N_{tot}  }\right) \left({\Delta N_{tot} \over \Delta t}\right)\Delta t
\eeq
Using a new matrix $A_{ij} = \overline{t}\overline{\tau} / t_i \tau_j$ for compact notation,
\beq
{\Delta N_i \over \Delta t} =  {\overline{t} \over t_i } { N_i(t) \over N_{tot}  }\left({\Delta N_{tot} \over \Delta t}\right) + \sum_j {N_i N_j \over N_{tot}} \left( A_{ij} F_{ij} - A_{ji} F_{ji} \right){1\over \overline{\tau}}.
\eeq
This is the Lotka-Volterra equation~\ref{eq:LV} again. The replicator dynamics equation~\ref{eq:replicator} can be obtained using the chain derivative:
\beq
{dN_i \over dt} = N_{tot} {dS_i \over dt} + S_i{dN_{tot} \over dt},
\eeq
which, if the $t_i$ do not differ significantly from the $\overline{t}$, reduces to
\beq
\Delta S_{i} =  \sum_j {1\over \overline{\tau}}S_iS_j \left( A_{ij} F_{ij} - A_{ji} F_{ji} \right) \Delta t, \quad A_{ij} = {\overline{t}\overline{\tau} \over t_i \tau_j}.
\label{eq:Shares}
\eeq
This replicator dynamics equation, in the binary form (imitation dynamics), has an antisymmetric exchange matrix $\alpha_{ij} = A_{ij}F_{ji}-A_{ji}F_{ji}$. 

I have thus described the rate of uptake, $\alpha_{ij} = A_{ij}F_{ji}-A_{ji}F_{ji}$, completely in terms of technology, market properties and choice of agents. These values can thus be compared to real diffusion timescales or used to parameterise diffusion dynamics in technology models in cases where such data is not available. Note that diffusion timescales are not related to the properties of individual technologies but, rather, to the properties of \emph{pairs} of technologies plus investor choices. This shows that diffusion timescales measured from historical time series are, in actuality, abstractions of many underlying processes that include decision-making, and should be understood to refer only to the decision context where they apply; they cannot be expected to represent other contexts where other choices might have been made.\footnote{Given, say, a different set of possibilities available to investors.}

As a final note, I ask, can this theory be put into a form that uses a multinomial logit instead, leading perhaps to a different form of the replicator equation? \ref{sect:AppC} presents in a demonstration that this is indeed approximately the case, where by grouping the binary logit terms, the classical multinomial logit is obtained. This naturally leads to a replicator equation of the classical form used in evolutionary theory,
\beq
\Delta S_i = S_i \left( \mathcal{F}_i(\vec{S}) - \overline{\mathcal{F}}(\vec{S}) \right) \Delta t,
\eeq
which is expressed in terms of the difference between the fitness $\mathcal{F}_i$, in the evolutionary theory sense, of technology $i$, to the average fitness $\overline{\mathcal{F}}$. This, however, is an approximation of the more accurate and practically usable binary system. The binary system is, effectively, probably the only way to correctly incorporate \emph{restricted access to technology and information}, which introduces a differentiation between options as seen by the agent, as opposed to a comparison with the `average' alternative.

\section{How to use this theory: Interpretation of the Lotka-Volterra scaling parameters \label{sect:interp}}
\subsection{Constraints and applicability of the Lotka-Volterra model \label{sect:conditions}}

The Lotka-Volterra model is a special case of the general model derived here from demography theory.  This section summarises the constraints under which this applies:
\begin{enumerate}
\item The birth and death functions have similar approximate widths in time,
\item The area under the birth function for technology $i$, $\Phi_i$, times the reinvestment fraction $R_i$, must be greater than one for a technology to be able to replicate itself.
\end{enumerate}
One then finds that, according to eq.~\ref{eq:Approx2}, $R_i \Phi_i$ determines the growth time constant in terms of the lifetime: $t_i = \tau_i/R_i \Phi_i$, where 
\beq
{1\over t_i} =  {R_i\over \tau_i}\int_0^\infty m_i(b) db,
\label{eq:scaling}
\eeq
the integral determining the total expected production by one unit of capital over its lifetime. Furthermore, since the productivity constant, after a possible delay of installation, is independent of age, then it can be further approximated to $1/t_i = R_i P_i \tau_i^K/\tau_i \simeq R_i P_i$. Thus we unsurprisingly find again that the rate of reinvestment $R_iP_i$ determines the magnitude of the fastest possible rate of growth of the production capacity $t_i^{-1}$ (see section \ref{sect:birth} and \ref{sect:AppA}). Thus in order for the industry to grow, such that $t_i < \tau_i$, one must have that $R_i \Phi_i > 1$, $\Phi_i$ representing the expected cumulative production of one installed unit of production capital during its lifetime.

From this, strict constraints can be determined that provide insight over which systems can be modelled using the Lotka-Volterra set of equations (LVEs) at the technology unit level:\footnote{LVE systems can be applied at other levels, e.g. firms. This may provide ways to deal with cases excluded here, requiring further research.} Furthermore, this model is also a deterministic one that does not include the process of generation of new technologies directly. The following are the factors that limit its predictive power:
\begin{enumerate}
\item If the lifetime of the production capital and that of the technology units it produces is very different, the LVEs are not appropriate when used at the unit level.\footnote{E.g. the mobile phone industry, in which phones have very short lifetimes, or infrastructure industries where the capital, e.g. houses, roads and bridges, have much longer lifetimes than the firms building them, potentially maintained forever.}
\item The producing firms must have an intended propensity towards expansion, and must reinvest enough profits to expand their production capacity, which will only decline if sales decline due to a lack of interest by investors/consumers (i.e. $R_i$ roughly constant). In a case where a firm has made a decision not to maintain a technology under production despite that it is profitable, the Lotka-Volterra model breaks down.
\item In evaluating the evolution of the market shares of firms for a particular market, the technology unit used in the Lotka-Volterra equation is crucial. This must be service producing technologies at the unit level (e.g. ovens, power plants, vehicles of different engine types, lighting devices, etc), not the service itself (e.g. a piece of bread, a kWh, a transport service, light) or long-lived infrastructure (e.g. houses or buildings, roads, airports, sets of transmission lines, bridges) likely to be maintained for lengths of time beyond foreseeable future.\footnote{This model does not apply at the firm level, as was done in \cite{Atkeson2007}.}
\item Innovation is not included directly in the demography/diffusion model, which does not predict the generation of new technologies, which could disrupt the diffusion of the other existing ones (e.g. the diffusion of fusion power). However, some aspects of innovation can, and should be, included through learning-by-doing cost reductions, which can easily be included in the decision model as described above (although the rates are uncertain). This approach narrows down possibilities for \emph{existing} technologies for the near future (e.g. during several life expectancies), which is itself quite robust, since it is well established that the formation of dominant designs is itself a lengthy process (e.g. the diffusion of fusion power). This limitation affects the realistic modelling time horizon.
\item Finally, this theory requires to be fully validated with historical data beyond existing empirical work, outside the scope of the present paper.
\end{enumerate}

\subsection{How to use this theory in real models}

Summarising this theory, for real models, both the industrial dynamics $A_{ij}$ and the decision-making $F_{ij}$ processes must be parameterised and used in the replicator equation for market shares $S_i$,
\beq
\Delta S_i = \sum_j S_i S_j \left( A_{ij}F_{ij} - A_{ji}F_{ji} \right) {\Delta t \over \overline{\tau}},
\eeq
requiring of course starting share values obtained from real-world data, and an absolute time scaling constant $\overline{\tau}$, the average life expectancy (see below).

\subsubsection*{Industrial dynamics $A_{ij}$}

Using the theory given above to parameterise a computational model of technology diffusion boils down to determining two parameters per technology: $t_i$ and $\tau_i$, 
\beq
\tau_i = \int_0^\infty \ell_i(a)da, \quad t_i = {\tau_i \over R_i \Phi_i} \simeq {1 \over R_i P_i}, \quad A_{ij} = {\overline{\tau}\overline{t} \over \tau_i t_j},
\eeq
with $\tau_i$ the life expectancy from the survival function, and $t_i$ the fastest possible growth rate in terms of the re-investment rate $R_i$ (in units re-invested per unit sold) and the productivity $P_i$ (in units produced per year), and $\overline{\tau}^{-1}, \overline{t}^{-1}$ are share weighted averages of the inverse of the time constants,
\beq
{1 \over \overline{t}} = \sum_i {S_i(t) \over t_i} \quad \text{and}\quad  {1 \over \overline{\tau}} = \sum_i {S_i(t) \over \tau_i}.
\eeq
Then this is put into the changeover timescale matrix for every possible pair of technologies $A_{ij}$.


Further simplifications are possible however. Since these timescales $t_i$ and $\tau_i$ only ever appear as ratios with their averages $\overline{t}$ and $\overline{\tau}$, the common scaling factors cancel out.\footnote{Note however that knowledge of the absolute value of $\overline{\tau}$, the absolute time scaling factor, appears on its own in the equation and is thus necessary.} Therefore, for instance, it is not the absolute value of $t_i$ that determines the rate of technology uptake, but its ratio with the average, in other words, \emph{how much faster is a technology able to fill gaps in the market in comparison to other technologies}. 

Furthermore, for technologies with long times of construction, for similar fractions of profit re-invested into production between technologies $R_i$, the productivity $P_i$ scales with the inverse of the time of construction, the time a firm has to wait before allocating its production capacity to new projects, and thus $\overline{t}/t_i = P_i/\overline{P}$ (e.g. power sector), equal to the ratio of the time of construction with the average, other parameters cancelling out. If times of construction are the same, however, but $R_i$ vary, then the $P_i$ cancel out and the ratio $R_i/\overline{R}$ must be used. Or it can also be that $t_i$ simply cancels out with $\overline{t}$ altogether and the $\tau_i$ broadly determine timescales of changeover (e.g. the car industry).

\subsubsection*{Technology choice $F_{ij}$}

$F_{ij}$ must be evaluated using a binary logit. This can be parameterised using measured cost distributions of sales, in which the diversity of past choices is represented. In the common case where small amounts of information on agents is available beyond cost distributions, the simplest approach is to parameterise the Gumbel or other type of distributions on the cost data, obtaining in this way a mean cost $C_i$ and a standard deviation $\sigma_i$ for every technology. Note that the shape of the distribution does not matter significantly in practice. The simple form of the binary logit can then be used for each technology category:
\beq
F_{ij} = {1 \over 1 + e^{\Delta C_{ij} / \sigma_{ij}}}, \quad \Delta C_{ij} = C_i - C_j, \quad \sigma_{ij} = \sqrt{\sigma_i^2 + \sigma_j^2}.
\eeq

\section{Conclusion}

This work demonstrates that the origin of the empirical observation of the applicability of the Lotka-Volterra or replicator dynamics models of competition dynamics to technology diffusion can be derived from demographic principles applied to technology. I have created an age structured model of technology demography, using life expectancies and birth rates which, given the right conditions, using an approximation, falls back onto the form of the well known empirical Lotka-Volterra and replicator dynamics models of competition. This procedure explains on the way the nature of the scaling parameters of the Lotka-Volterra equation, the timescales of technology diffusion, in terms of \emph{survival} properties of technology and industrial dynamics stemming from investment. 

The calculation presented however generates more insight than the simple correspondence of the Lotka-Volterra system to demography. While every previous quantitative use of the Lotka-Volterra system for modelling technology diffusion has remained empirical and without clear explanation of its parameterisation, the calculation presented here explains \emph{why} the Lotka-Volterra actually describes well systems of competing technologies at all. It moreover clarifies under which conditions it applies. Meanwhile, this paper gives meaning of the timescales of technology population dynamics measured empirically. 

By clarifying the meaning of the scaling constants of the Lotka-Volterra model, this theory enables its use with a method for its parameterisation without prior empirical measurement, the latter difficult to achieve in cases where only small amounts of data are available. This tends to occur precisely in the cases of interest, namely when exploring the diffusion potential of new technologies under different assumptions over the market and policy environment. This theory thus enables to build models of technology forecasting based on $S$-shaped diffusion curves and to parameterise them using known properties of the technologies and those of their respective production industries. This method, as used for instance in earlier work \citep{Mercure2012, Mercure2014}, can in principle replace the optimisation algorithms in mainstream models which have little theoretical foundation. This opens many possibilities for modelling future technology pathways, for instance for analysing the impacts of policy supporting the diffusion of low carbon technology.


\section*{Acknowledgements}

I would like to acknowledge the students who have been attending the Energy Systems Modelling seminars held at 4CMR in 2013, in particular A.~Lam, P. Salas and E.~Oughton, with whom the discussions on technology and evolutionary economics led to the development and enhancement of this theory. I would furthermore like to thank participants to the International Conference on Sustainability Transition 2013 for providing valuable feedback, in particular K.~Safarzynska for critical comments, as well as two anonymous referees for their recommendations. This research was funded by the UK Engineering and Physical Sciences Research Council, fellowship number EP/K007254/1.

\setcounter{section}{0}
\setcounter{figure}{0}
\renewcommand*\thesection{Appendix \Alph{section}}
\renewcommand*\thefigure{\alph{section}.\arabic{figure}}

\section{: Growth under full employment \label{sect:AppA}}

In a situation where an innovation was to follow a path free of competitors and grow as fast as its industry can produce it, it would follow the fastest possible rate $t_i^{-1}$ of exponential growth determined by:
\beq
1 = \int_0^\infty R_i e^{-b/t_i} \ell_i^K(b) P_i^K(b) db.
\label{eq:transcen}
\eeq
This is a transcendental equation that can only be solved numerically. As demonstrated by \cite{Lotka1911} \citep[see][ for a clearer derivation]{Kot2001}, it has only one real solution, all others being complex of the form $u\pm iv$ which give rise to oscillatory behaviour in the real part. The non-oscillatory real solution, an exponential, can be approximated if simple forms are taken for $\ell(b)$ and $P(b)$ (see fig.~\ref{fig:SurvivalFunction}, \emph{right}):

\vspace{.3cm}
	\begin{minipage}[l]{0.4\columnwidth}
		\begin{center}
			\includegraphics[width = .7\columnwidth]{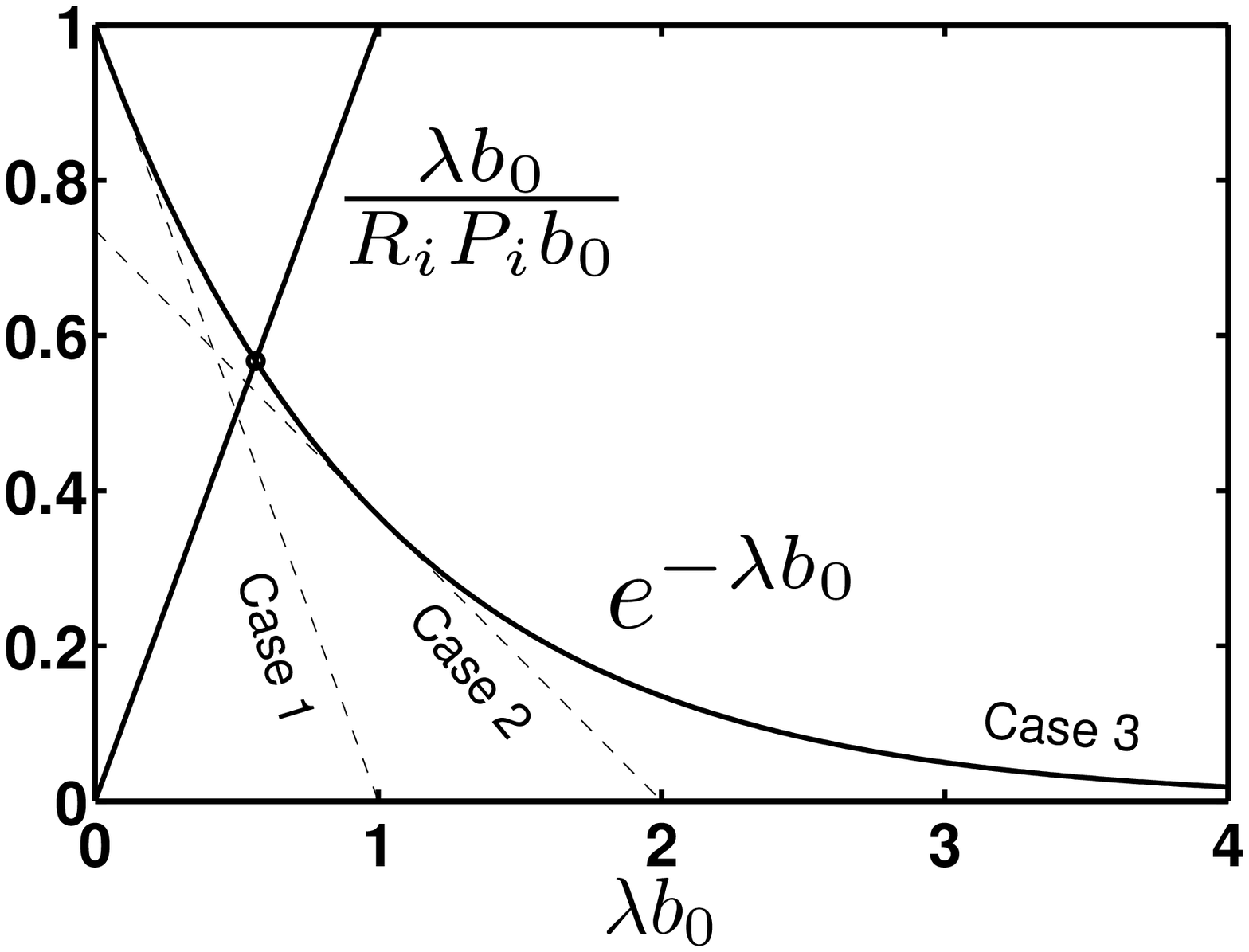}
		\end{center}
	\end{minipage}
	\begin{minipage}[l]{0.6\columnwidth}
		\beq
		\ell_i^K(b) \simeq e^{-b/\tau_i^K}, \quad P(b) = \left\{ \begin{array}{rl} 0, &\quad b < b_0 \nn\\ P_i, &\quad b \geq b_0\end{array}\right.
		\eeq
		\beq
		\Rightarrow {\lambda b_0 \over (R_i P_i b_0)} = e^{-\lambda b_0} \quad \text{with} \, \lambda = {{1\over t_i} + {1 \over \tau_i^K}},\nn\\
		\eeq
	\end{minipage}
	\label{fig:Approx}
where $b_0$ is the time between investment and construction, $P_i$ is a production rate and therefore $(R_iP_i)^{-1}$ is the rate of expansion of production capacity (in inverse years), and $\tau_i^K$ is the timescale of capital depreciation, its life expectancy (and therefore we always have $\tau_i^K >> b_0$). To first order, one can find which is the dominant of these timescales in particular situations, using limits for the dimensionless parameter $R_i P_i b_0$ (which determines the slope of the linear left hand side of the equation, see graph).

\vspace{8pt}
\textbf{Case 1: $(R_iP_i b_0)$ is small $\Rightarrow$ $\lambda b_0$ is small} \\
We perform a Taylor expansion around $\lambda b_0 = 0$, 
\beq
\lambda = R_iP_i \left( 1 - b_0 \lambda - {b_0^2\lambda^2 \over 2} + ... \right)\quad \Rightarrow \quad t_i \simeq \left[ {1 \over (R_iP_i)^{-1} + b_0} - {1 \over \tau_i^K}\right]^{-1}
\eeq
Since $R_iP_i$ cannot be a very small quantity, it is most likely that $b_0 << (R_iP_i)^{-1}$, for instance with small technologies that are ready to use as they come out of the factory (e.g. vehicles). And since $R_iP_i >> 1/\tau_i^K$, then $t_i \simeq (R_i P_i)^{-1}$. In this case the rate of production is constrained by the rate of re-investment into production capacity, which depends on the rate of production of technology units but not on the time of construction of production capacity. For large systems (e.g. power plants, wind turbines, infrastructure), the time of construction may be long (i.e. several years), constraining money flows used for firm expansions. For small modular technologies (e.g. electronics), the time of production is short and other timescales dominate the time `bottleneck'.

\vspace{8pt}
\textbf{Case 2: $(R_iP_i b_0)$ is of order 1 $\Rightarrow$ $\lambda b_0 \simeq 1$} \\
We perform a Taylor expansion around $\lambda b_0 = 1$, 
\beq
\lambda = R_iP_i e^{-1} \left( 1 - (b_0 \lambda-1) - ... \right)\quad \Rightarrow \quad t_i \simeq \left[ {1 \over (eR_iP_i)^{-1} + b_0} - {1 \over \tau_i^K}\right]^{-1}
\eeq
If $b_0 << R_i^{-1}P_i^{-1}$ then $t = e R_i P_i$ and the limiting timescale is again the re-investment rate. However, if $b_0 >> R_i^{-1}P_i^{-1}$ then $t = b_0$ and the rate of growth is limited by the rate of completion of capital installation. For example, for technologies with complex production capital structures with a long time of for installation with no income constrains the rate of return, the dominant \emph{bottleneck} timescale is $b_0$, a situation where a firm must wait for expansion projects to come to completion and income to be brought in before launching itself into further expansions. 

\vspace{8pt}
\textbf{Case 3: $(R_iP_i b_0)$ is large $\Rightarrow$ $\lambda b_0$ is large} \\
No Taylor expansion is possible. However if $R_iP_i e^{-\lambda b_0} \rightarrow 1/\tau_i^K$, then
\beq
{1 \over t_i} = R_i P_i e^{-\lambda b_0} - {1 \over \tau_i^K} \simeq \rightarrow 0
\eeq
and the timescale of expansion diverges. This corresponds to a case where a firm struggles with its cash flows to maintain its production capacity. Beyond this the timescale can also become negative, where a firm scales down its activities.

Thus in many cases one of the three timescales dominates, the bottleneck timescale. In other cases, if two timescales are similar, $t_i$ must be calculated numerically using eq.~\ref{eq:transcen}.

These three cases however only occur if consumers are ready to buy \emph{all} that this particular industry is able to produce, and consequently its growth is limited by its ability to expand. If, however, the demand grows more slowly than these maximal rates, then the demand constrains the rate of growth, a demand-led case. Furthermore, if consumers have a choice of products and competition occurs, then the rate of growth is further constrained and a model of competition must be derived, as done in section~\ref{sect:model2}.

\section{: A binary logit choice model \label{sect:AppB}}

A model of choice is constructed here using a pairwise comparison, which will be performed for all possible pairs in order to rank exhaustively consumer preferences, the latter being distributed. I use for this generalised cost distribution of sales obtained from recent sales data. This calculation is the basis of discrete choice theory adapted to the purposes of this work, and more information can be obtained in \cite{Ben-Akiva1985} and \cite{Domencich1975}.

I assume two distributions for the relative numbers of situations where agents, stating their individual preference between technologies $i$ and $j$, face different situations and state different choices. By counting how many agents prefer which technology in each pair, one can determine what the probabilities of preferences between these two technologies are for future situations where choices are to be made (e.g. 70\% of agents choose $i$ and 30\% $j$). It does not mean however that when the time comes to invest or purchase, these are the choices that would be made, since depending on the state of diffusion of these technologies, they might not necessarily be available to every agent. By going through an exhaustive list of pairwise preferences, final choices can be determined.

I denote these (normalised) distributions $f(C,C_i,\sigma_i)dC = f_i(C-C_i)dC$ and $f(C,C_j,\sigma_j)dC = f_j(C-C_j)dC$, where $C_i,C_j$ are the means and $\sigma_i,\sigma_j$ are the standard deviations for technologies $i$ and $j$. These distributions can be of any kind, but they require to have a single well defined maximum and variance (e.g. they cannot have two maxima\footnote{In which case we would need to subdivide such a technology category into two.}). I can then evaluate the probability of choosing $i$ over $j$ using the following. First, I calculate the probability of choosing $i$ in all cases where $j$ has an arbitrary cost $C$. The central assumption here is that the fraction of agents for whom the generalised cost of $j$ is $C$ and for whom the cost for $i$ is lower than $C$ will choose technology $i$ over $j$ if given a choice, and this fraction is equal to the cumulative probability distribution $F_i(C-C_i)$. But this situation occurs a fraction $f_j(C-C_j)$ of the time, giving a total probability
\beq
P(C_i < C | C_j = C) = F_i(C-C_i) f_j(C-C_j)dC,
\eeq
while the converse is
\beq
P(C_j < C | C_i = C) = F_j(C-C_j) f_i(C-C_i)dC.
\eeq
In order to evaluate how often the cost of technology $i$ is lower than that of technology $j$, and the converse, a sum over all possible values of $C$ must be taken. For simplicity, I use as variables $C' = C-C_j$ and $C'' = C - C_i$, with the mean cost difference $\Delta C = C_i - C_j$:
\beq
F_{ij}(\Delta C) = P(C_i < C_j) = \int_{-\infty}^{+\infty}  F_i(C'-\Delta C) f_j(C')dC',\nn
\eeq
\beq
F_{ji}(\Delta C) = 1- F_{ij} = P(C_j < C_i) = \int_{-\infty}^{+\infty} F_j(C''+\Delta C) f_i(C'')dC''.
\eeq
This appears difficult without further knowledge of the distribution type, however it is possible to take a derivative with respect to $\Delta C$, which makes the integral a convolution of the two distributions
\beq
{d F_{ij} \over d\Delta C}  = -\int_{-\infty}^{+\infty} f_i(C'-\Delta C) f_j(C')dC' = - f_{ij}(\Delta C) 
\eeq
\beq
= \int_{\infty}^{-\infty} f_i(C'') f_j(C''+\Delta C)dC'' = - f_{ji}(-\Delta C) = {d F_{ji} \over d\Delta C}.
\eeq
This convolution yields a new distribution $f_{ij}(\Delta C) d\Delta C$ of which the standard deviation is $\sigma_{ij} = \sqrt{\sigma_i^2 + \sigma_j^2}$. This is the probability distribution of technology switching in terms of $\Delta C$. The convolution having been computed, this distribution can be integrated again as a function of $\Delta C$ to yield a cumulative probability distribution that the cost of technology $i$ is less than that of $j$ (and conversely):
\beq
F_{ij}(\Delta C)  = \int_{-\infty}^{+\infty} f_{ij}(\Delta C)d\Delta C = 1 - \int_{-\infty}^{+\infty} f_{ji}(\Delta C)d\Delta C = 1-F_{ji}(\Delta C).
\eeq
Thus given a choice between technologies $i$ and $j$, the fraction $F_{ij}$ of agents tends to choose technology $i$ and the fraction $F_{ji}$ chooses $j$, these fractions being functions of the generalised cost difference, and this cumulative choice function has a width that follows the sum of the squares $\sigma_{ij} = \sqrt{\sigma_i^2 + \sigma_j^2}$. Note that this calculation is independent of probability distribution type; however $F_{ij}(\Delta C)$ should have roughly the shape of a `smooth' step function, its `smoothness' determined roughly by the widths of \emph{both} cost distributions.

In discrete choice theory, the Gumbel distribution is often used, $f_i = e^{-e^{-(C-C_i)/\sigma_i}}$, and the result of the convolution of two Gumbel distributions is a logistic distribution of the average cost difference $\Delta C_{ij}$ relative to the root mean square width $\sigma_{ij}$:
\beq
f_i = e^{-e^{-(C-C_i)/\sigma_i}}, \quad F_{ij} = \int_{-\infty}^\infty \left( f_i \ast f_j \right) d\Delta C = {1 \over 1 + e^{\Delta C_{ij}/\sigma_{ij}}}.
\eeq

\section{: From the binary to the multinomial logit in the replicator equation \label{sect:AppC}}

The derivation of the binary logit in \ref{sect:AppB} gives a relationship between the cost probability distributions and the cumulative distribution of choice between two options $f_i$ and $f_j$ (with parameters $C_i,\sigma_i$ and $C_j,\sigma_j$), as a convolution, consistent with \cite{Domencich1975}. The probability of cost of option $i$ being less than the cost of option $j$ is
\beq
P(C_i <  C_j) = {d F_{ij} \over d\Delta C_{ij}}  = -\int_{-\infty}^{+\infty} f_i(C'-\Delta C_{ij}) f_j(C')dC' = f_i \ast f_j,
\eeq
the star denoting a convolution. The binary form of the replicator dynamics equation requires summing the result of binary choices, however distributions can be first grouped and afterwards convolved:
\beq
\sum_j A_{ij}F_{ij} S_j = \int_{-\infty}^{\infty} \left(f_i \ast \sum_j A_{ij}S_j f_j \right) d\Delta C_{ij}.
\label{eq:gi}
\eeq
In this picture, each cost distribution of possible alternatives $f_j$ is weighted by the factor $A_{ij}S_j$ which involves shares and changeover timescales. This weighted sum of distributions results in a composite distribution with new mean and standard deviation parameters $\overline{C}$ and $\overline{\sigma}$.\footnote{$\overline{C}$ is the weighted average, while $\overline{\sigma}^2$ is the weighted sum of the square of the standard deviations.}, which cannot be expressed analytically in any simpler form. The convolution corresponds roughly to the probability that the cost of option $i$ is less than the cost of the `average' alternative (with average cost $\overline{C}$) weighted by the frequency of occurrence of these choices, $P(C < \overline{C} | C_i = C)$. 

As an approximation, I replace this `average' probability distribution with that of an arbitrary cost value $C$ being lower than the minimum of all available alternatives simultaneously. This corresponds to the product of the individual distributions,
\beq
P\left(C < \min[ C_1, C_2, C_3 ... C_n] \right) = P(C < C_1)P(C < C_2) ... P(C < C_n).\nn
\eeq
When the weighting of these choices is equal, if each of these probability functions are Gumbel, then the result of this is also a Gumbel distribution, of cost parameter proportional to $ \ln \sum_j e^{-C_j / \sigma_j}$, a classic result of discrete choice theory when deriving the multinomial logit \citep[][p.~105]{Ben-Akiva1985}. Equal weighting in the multinomial logit corresponds to perfect access to information and technology options by all agents. In the theory presented here, each agent has access to a different set of choices, which, when correctly weighted, is
\beq
P(C < C_1)^{A_{i1}S_1}P(C < C_2)^{A_{i2}S_2} ... P(C < C_n)^{A_{in}S_n},\nn
\eeq
\beq
P\left(C < \min[ C_1, C_2, C_3 ... C_n] | C_i = C \right) = F_i(C_i-C) \prod_j P(C < C_j)^{A_{ij}S_j}.
\label{eq:min}
\eeq
The weighted (representative alternative) cost parameter is instead
\beq
\hat{C} = \hat{\sigma} \ln \sum_j A_{ij}S_j e^{-C_j \over \sigma_j} .
\eeq
This unequal weighting of alternatives is generally overlooked in discrete choice theory, but crucial when exploring the diffusion of technology since part of the dynamics stem from restricted access to options in early states of diffusion.  Following \ref{sect:AppB}, the convolution becomes
\beq
\int_{-\infty}^\infty \left( f_i \ast \hat{f} \right) d\Delta C_i = {1 \over 1 + e^{A_{ii}{C_i \over \sigma_i} - \ln \left( \sum_j A_{ij}S_j e^{-C_j / \sigma_j} \right)}} = {A_{ii} e^{-C_i /\sigma_i} \over  \sum_j A_{ij}S_j e^{-C_j / \sigma_j} }.
\eeq
This is the multinomial logit weighted by $A_{ij}S_j$, i.e. adjusted for restricted access to alternatives. I now define the fitness in the evolutionary theory sense,
\beq
\mathcal{F}_i = {{\overline{t} \over t_i} e^{-C_i /\sigma_i} \over  \sum_j {\overline{t} \over t_j} S_j e^{-C_j / \sigma_j} } - {{\overline{\tau} \over \tau_i} e^{-C_i /\sigma_i} \over  \sum_j {\overline{\tau} \over \tau_j} S_j e^{-C_j / \sigma_j} } 
\eeq
This is the fitness of a technology to capture the market, a growth minus a survival term.\footnote{Producing faster (smaller $t_i$) increases competitiveness and therefore the fitness, while surviving for longer (larger $\tau_i$) decreases vulnerability to changes and thus also improves the fitness.} The average fitness is then
\beq
\overline{\mathcal{F}} = \sum_j S_j \mathcal{F}_j = 1 - 1 = 0.
\eeq
Thus the replicator dynamics equation (eq.~\ref{eq:Shares}) can in fact be written as
\beq
\Delta S_i = S_i \left( \mathcal{F}_i - \overline{\mathcal{F}} \right) \Delta t.
\eeq
This is the classical replicator dynamics equation in general evolutionary theory \citep[e.g.][]{Hofbauer1998}, where the ability of a proponent option or biological specie to capture market or space is proportional to the difference of its fitness to the average fitness.

This transformation however has required an approximation which is a simplification of the distribution of alternatives. This is a useful simplification for the sake of exposition, but leads to a less accurate form of the replicator equation and associated market response. This is due to leaving out, at eq.~\ref{eq:min}, some of the details of the restricted access to technology and information, as well as the complexity emerging from interactions.\footnote{Replacing all alternatives by a single `representative' alternative.} The binary form essentially maintains the information as to which options are seen by which agent in aggregate. It is however heavier computationally since it involves pairwise comparisons, which scales as $n^2/2 - n$ for the binary form, compared to $n$ for the multinomial form ($n$ the number of options).

\section*{References}

\bibliographystyle{elsarticle-harv}
\bibliography{../../../../CamRefs}

\end{document}